\documentclass[doublecol]{epl2} 
\usepackage[british]{babel}
\usepackage[usenames,dvipsnames]{xcolor}
\usepackage{amsmath,amssymb}
\usepackage[normalem]{ulem}
\usepackage{mcite}
\usepackage{url}

\newcommand{\ch}[1]{\textcolor{black}{#1}}
 
\title{Polariton condensation with saturable molecules dressed by
  vibrational modes}

\abstract{ Polaritons, mixed light--matter quasiparticles, undergo a
  transition to a condensed, macroscopically coherent state at low
  temperatures or high densities.  \ch{Recent experiments show that coupling light to organic molecules inside a microcavity allows condensation at room temperature.  The molecules act as
  saturable absorbers with transitions dressed by molecular
  vibrational modes.  Motivated by this we calculate the phase diagram
  and spectrum of a modified Tavis--Cummings model, describing vibrationally dressed two-level systems, coupled
  to a cavity mode.  Coupling to vibrational modes can induce
  re-entrance, i.e.\ a normal-condensed-normal sequence with decreasing
  temperature and can
  drive the transition first order.  }}

\author{Justyna A. {\'C}wik\inst{1} \and Sahinur Reja\inst{2} \and
  Peter B. Littlewood\inst{3,4} \and Jonathan Keeling\inst{1}}
\shortauthor{Justyna A. {\'C}wik \etal}

\institute{                    
  \inst{1} SUPA, School of Physics and Astronomy, University of St Andrews, St Andrews KY16 9SS, United Kingdom\\
  \inst{2} Cavendish Laboratory, University of Cambridge, Cambridge CB3 0HE, United Kingdom\\
  \inst{3} Physical Science and Engineering,
  Argonne National Laboratory,  9700 S. Cass.\ Av., Argonne, IL 60439, USA\\
  \inst{4} James Franck Institute and Department of Physics,
  University of Chicago, IL 60637, USA
}

\begin{document}

\maketitle

\section{Introduction}
\ch{Microcavity polaritons (superpositions of photons and excitons) are
bosonic quasiparticles which can form a
Bose--Einstein condensate (BEC)~\cite{Kasprzak2006,snoke07science}.
Experimentally,  above a critical density or below a
critical temperature  polaritons accumulate in low energy modes,
accompanied by enhanced spatial and temporal coherence.  Such
coherence  naturally relates to lasing, but differs in that the occupied states are polaritons, not cavity photons, and
the coherence results from stimulated scattering, not stimulated
emission~\cite{Deng2010,Carusotto2013}.}  At low densities the critical
temperature for condensation goes as $k_B T_c \sim \hbar^2 \rho/m$
with $\rho$ and $m$ the density and the mass of polaritons
respectively.  \ch{At higher densities, when $k_B T_c$
reaches the scale of the light--matter coupling, there is a crossover
to an almost density independent form~\cite{Keeling2007a}.}
Consequently, in order to reach higher temperatures than the $20-100$K
attained in CdTe~\cite{Kasprzak2006} and GaAs~\cite{snoke07science},
one requires materials with larger light--matter coupling, such as GaN
and ZnO~\cite{Zamfirescu2002,Christopoulos2007,Li2013}. \ch{Alternatively,
one may replace the Wannier excitons in inorganic semiconductors with
electronic excitations in organic molecules, which may have large
oscillator strengths,   allowing room temperature condensation.}

\ch{Recent experiments on organic based microcavity polaritons have
  explored a wide variety of organic materials. These include
  molecular crystals of anthracene~\cite{Kena-Cohen2008,Forrest2010},
  molecular aggregates coupled by F{\"o}rster transfer, \textit{e.g.}\
  J-aggregates of cyanine dyes~\cite{Lidzey1998,Tischler2005} and
  amorphous molecular structures of conjugated
  polymers~\cite{Plumhof2013}. Several of these systems have shown
  condensation or lasing of polaritons at room temperature: Polariton
  lasing has been reported with molecular crystals of
  anthracene~\cite{Forrest2010}, and polariton lasing using
  J-aggregates, where a separate organic dye acts as a gain medium has
  been seen~\cite{intralasing2013}.  Very recently, experiments on
  amorphous materials have shown interacting polariton
  condensates~\cite{daskalakis14}, and the formation of a thermalised
  and an interacting polariton BEC~\cite{Plumhof2013}.}

\ch{The materials used in these experiments differ in their structure, and especially in the mechanisms
for electronic excitation transfer between and within molecules.}
However, as discussed below, these differences have a reduced
significance in the presence of strong light--matter coupling.  
\ch{As well as the obvious significance of room temperature BEC, understanding polariton condensation in these organic materials may facilitate lower lasing thresholds enabling electrically pumped organic lasers~\cite{Samuel2007}.}
\ch{The theory of excitons and polaritons in molecular crystals has a long
history~\cite{Davydov1971, AgranovichRussian, Agranovich2009a}.  Recent theoretical work on
polariton condensation has generally modelled the system as a weakly
interacting gas of polaritons derived from a model of
saturable absorbers~\cite{Litinskaya2008}.  Rate
equations based on such models have been used to calculate the
luminescence spectrum~\cite{Fontanesi2009,Mazza2009} and relaxation
processes~\cite{Michetti2009,Mazza2013a}.} The effects of disorder on the
spectrum~\cite{Litinskaya2004,Litinskaya2006} have also been
considered.  Such theories well describe the very low density regime.
The approach we will describe below, starting from two-level systems,
encompasses also higher densities.

\ch{We focus on molecular crystals, consisting of many separate
saturable optical absorbers.  In the absence of an optical cavity, hopping of excitations between molecules is crucial in determining the band structure and polarisation properties of Frenkel excitons~\cite{Davydov1971, AgranovichRussian, Agranovich2009a}.
However, when strongly coupled to an optical cavity, the rate of
hopping between molecules is dwarfed by cavity--photon mediated
transport. The
effective exciton mass due to the exciton hopping is four
orders of magnitude larger than the photon mass, and the polariton
splitting is at least an order of magnitude larger than exciton
bandwidth.} Consequently, exciton hopping can generally be neglected
when considering the thermodynamics of polaritons.

Following these considerations, we study a model of two-level systems,
describing localised electronic excitations of the molecules, coupled
to a common photon mode.  This is a variant of the
Dicke~\cite{dicke54} or Tavis--Cummings~\cite{Tavis:TC} model. \ch{Hepp
and Lieb~\cite{Hepp:Super} showed that in the canonical ensemble above
a critical light--matter coupling strength, the system undergoes a
continuous phase transition from a normal to superradiant state.}
\ch{The critical temperature of this transition can be suppressed to
  $T=0$, producing the quantum phase transition much discussed for
  cavity and circuit QED~\cite{Emary2003,Dimer2007,Nataf2010}. It has,
  though, long been thought that the phase transition of this model in
  the canonical ensemble is forbidden~\cite{Rzazewski1975}. Recently,
  however, this debate has been re-opened\cite{Nataf2010,Viehmann2011,
    Ciuti2012} suggesting that the transition in the canonical
  ensemble may in fact be possible~\cite{Vukics2012a}. In this paper
  we focus on the grand canonical ensemble where this issue does not
  arise~\cite{Eastham:Localized}.
In this ensemble
the phase transition is identical to the BEC
transition~\cite{Eastham:Localized,Eastham2001,Keeling2007a}, with the
superradiant state corresponding to the condensate.}

\ch{Work on kinetics of polariton relaxation~\cite{Michetti2009,Mazza2013a} has
shown the crucial role of the local vibrational modes in energy
relaxation: Polariton relaxation is most efficient when the polariton splitting is resonant with the vibrational frequency. The importance of strong coupling to vibrational modes has
also been  recognised in contexts such as energy transfer in
light harvesting complexes~\cite{Engel2007,
  *Panitchayangkoon2010,China}.}  In particular, these works show that
such vibrational modes cannot simply be regarded as Markovian baths
leading to dephasing or dissipation.
\ch{In order to consider such effects within the context of saturable optical absorbers, we need to augment the Tavis--Cummings model by introducing an additional
feature: Coupling between electronic
excitations and local vibrational modes of the molecules.}

In this Letter, we consider the effects of coupling between the
two-level systems and local vibrational modes on collective behaviour
within the Tavis--Cummings model.  Current experiments are far from
equilibrium due to the combination of loss of photons, external
incoherent pumping and the dephasing of vibrational modes caused by
coupling to other molecules.  Coupling between electronic
  excitations and vibrational modes can act as a route to relaxation
  and thermalisation, but as we will show, this is not its only
  r\^ole.  The driven--dissipative Tavis--Cummings model has been
used to explore the crossover between polariton condensation and
``textbook'' lasing~\cite{Szymanska2006,Keeling2013}, and so the model
we present provides a basis to address similar questions in the
presence of vibrational modes. However, in this Letter we focus
  on first providing a firm foundation in thermal equilibrium, as a
  reference to which the out-of-equilibrium physics can be compared.
We show how coupling to vibrational modes modifies the phase diagram;
such modifications are greatest when there is strong coupling to soft
vibrational modes, so we focus on such a regime.  To explain this
behaviour we consider the excitation spectrum and discuss how the
typical BEC scenario of condensation when the chemical potential hits
the polariton spectrum is modified by the presence of multiple
vibrational sidebands.  Finally, we consider the limit of very large
coupling to vibrational modes, and explain how this can drive the
phase transition first order.


\section{Model}

The model we study generalises the Tavis--Cummings
model~\cite{Tavis:TC}, which describes $N$ two-level systems (electronic states of molecules)
coupled to a single photon mode in the microcavity.  To this we add a
coupling between two-level systems and vibrational modes of the
molecules.  We thus have (setting $\hbar=1$ throughout):
\begin{multline}
  \hat{H}-\mu \hat{L}=\tilde{\omega}_{c} \hat{\psi}^{\dagger}\hat{\psi} + 
  \sum_{n=1}^{N}
  \left[
  \frac{\tilde{\epsilon}}{2}\sigma^{z}_{n}+g\left(\sigma^{+}_{n}\hat{\psi}+
  \hat{\psi}^{\dagger}\sigma^{-}_{n}\right)+ \right. \\
  +\left. \Omega \hat{a}^{\dagger}_{n}\hat{a}_{n}+\frac{\Omega
    \sqrt{S}}{2}\sigma_{n}^{z}\left(\hat{a}_{n}+\hat{a}^{\dagger}_{n}\right) \right].
\label{eqn:H_with_polaron}
\end{multline}
\ch{Here $\hat{L}=\hat{\psi}^\dagger \hat{\psi} + \sum_{n} \sigma
  ^{z}_{n}/2$ is the total number operator, $\hat{\psi}^\dagger$ is
the creation operator for a cavity photon, $\tilde{\omega}_c =
\omega_c - \mu$, where $\omega_c$ is the photon 
frequency.}  Two-level systems are described by the Pauli matrices
$\sigma^i_n$, $\tilde{\epsilon} = \epsilon - \mu$ where $\epsilon$ is
the bare optical transition frequency of the molecules, and the bare
molecule--photon coupling strength is denoted $g$.  \ch{Vibrational
excitations of a molecule, with frequency $\Omega$, are created by the
operators $\hat a^\dagger_n$. The coupling of these excitations to the
electronic state of the molecule is $\Omega\sqrt{S}$, where
 the Huang--Rhys parameter
$S$ quantifies the average number of vibrational
  excitations emitted or absorbed in an electronic transition.  The
characteristic polariton splitting is determined  by 
$g\sqrt{N}$.  (NB, as discussed later, $g\sqrt{N}$ is only
\emph{equal} to the polariton splitting in the absence of coupling to
vibrational modes.)  The chemical potential $\mu$ controls the total
excitation density, $\langle L \rangle$.
Eq.~(\ref{eqn:H_with_polaron}) describes a cavity single mode, while a
  planar microcavity supports multiple transverse cavity modes with different
  in-plane momenta.  Neglecting these other modes is
  equivalent to neglecting the depletion of the condensate by
  long wavelength fluctuations, \textit{i.e.}\ considering the
  mean--field (MF) theory. In~\cite{Keeling2007a} it was shown that
  these fluctuations are only relevant at extremely low densities and
  that MF theory, as we consider in this manuscript, is otherwise
  accurate.}

\ch{We next discuss the parameters we use when presenting numerical
results.}  In the following we measure energies in units of a
characteristic scale $g_0 \sqrt{N}$, corresponding to a typical
polariton splitting.  In some cases it is useful to show evolution of
phase boundaries with $g$, in which case we plot as a function of
$g/g_0$.  \ch{For the energies $\omega_{c},~\epsilon$, the physically important quantity is the photon--exciton detuning, $\Delta= \omega_c - \epsilon$.
We consider a case where the cavity frequency is detuned above the molecular transition and we choose $ \Delta = 2 g_0\sqrt{N}$.}  We choose
this detuning for two reasons.  Firstly, a thermal equilibrium BEC of
polaritons in inorganic materials~\cite{Kasprzak2006} required $\Delta
>0$, as this increases the excitonic fraction, and hence the
scattering and thermalisation rate of the
polaritons~\cite{doan05:prb,Kasprzak2008b}.  Secondly, the
Tavis--Cummings model can show multiple normal--superradiant phase
transitions~\cite{Eastham2001} when $\Delta > 0$. This is
linked~\cite{Schmidt2013} to the ``Mott-lobes'' in the
Jaynes--Cummings--Hubbard model~\cite{Greentree2006}.  We
  explore whether such Mott lobes survive coupling to vibrational
  modes.

  \ch{The parameters $S$ and $\Omega/g_0\sqrt{N}$ control the
  effects of the vibrational modes. To clearly observe the effects of
  vibrational dressing it is necessary to consider relatively large
  Huang--Rhys factors, corresponding to  ``ultrastrong''
  coupling to vibrational modes.  We  present results for both
  $S=2$ and $S=6$.  While these values are quite large for organic
  emitters (for anthracene~\cite{Malagoli2004} $S=0.182$), values such
  as $S=3.3$ have been seen for LO phonons in carbon
  nanotubes~\cite{Leturcq2009}.   Some features of
  vibrational dressing do survive for small $S$, though in a subdued
  manner, particularly the behaviour of the excitation spectrum.}
  Regarding $\Omega$, distinct behaviour occurs for soft modes,
  $\Omega \ll g_0\sqrt{N}$, {\it vs} stiff modes, $\Omega \sim
  g_0\sqrt{N}$.  We show that a soft mode is required for a re-entrant
  phase boundary, while the first order transition requires a stiffer
  vibrational mode.  We thus present results for
  $\Omega/g_0\sqrt{N}=0.05, 0.5$.  The latter value is comparable to
  that for anthracene~\cite{Malagoli2004} $\Omega=42$meV measured in
  units of the polariton splitting of Ref.~\cite{Kena-Cohen2008}. Such
  stiff modes arise due to the $\pi$-bonded carbon rings.

\section{Phase diagrams \& Re-entrance}

\ch{The Hamiltonian~(\ref{eqn:H_with_polaron}), supports two distinct
  phases: normal and condensed.  The order parameter distinguishing
  these phases is the expectation of the photon field $\langle
  \hat{\psi} \rangle$. In the following, we shall define a
  rescaled order parameter $\lambda=\langle \hat{\psi}
  \rangle/\sqrt{N}$. In the normal phase $\lambda=0$, whereas in the
  condensed phase there is a macroscopic expectation of the photon
  field so $\lambda \neq 0$.  In the various phase diagrams, we plot a
  colour map of the rescaled order parameter, $\lambda$.}

\ch{The equilibrium phase diagram can be calculated within a MF treatment of the photon
field, known to be exact~\cite{Eastham2001} in the thermodynamic limit
$N \to \infty$, $g \sqrt{N} \to$ const.}
\ch{The MF theory yields the self-consistency condition $\tilde{\omega}_c
\lambda = - g \sqrt{N} \langle \sigma^- \rangle$, where
the polarisation $\langle \sigma^-
\rangle$ is found by exactly diagonalising the on-site problem 
\ch{
\begin{equation}
h = \left[\tilde{\epsilon} + \Omega \sqrt{S}\left(\hat a+\hat a^\dagger \right)\right] \frac{\sigma^z}{2} +g\sqrt{N} \left(\lambda\sigma^{+}+ \text{H.c.}\right)+ \Omega
\hat{a}^{\dagger}\hat{a}
\label{eqn:exactDiag}
\end{equation} 
}numerically, whilst truncating the maximum number
of vibrational excitations, $n_{max}$, at $n_{max} \gg S$, and thermally
populating the resulting eigenstates.}  Anticipating possible first
order transitions, one must also compare the free energies of the
normal ($\lambda =0$) and condensed ($\lambda \neq 0$) solutions to
determine the global minimum free energy.

\begin{figure}[h]
\centering
\includegraphics[width=3.4in]{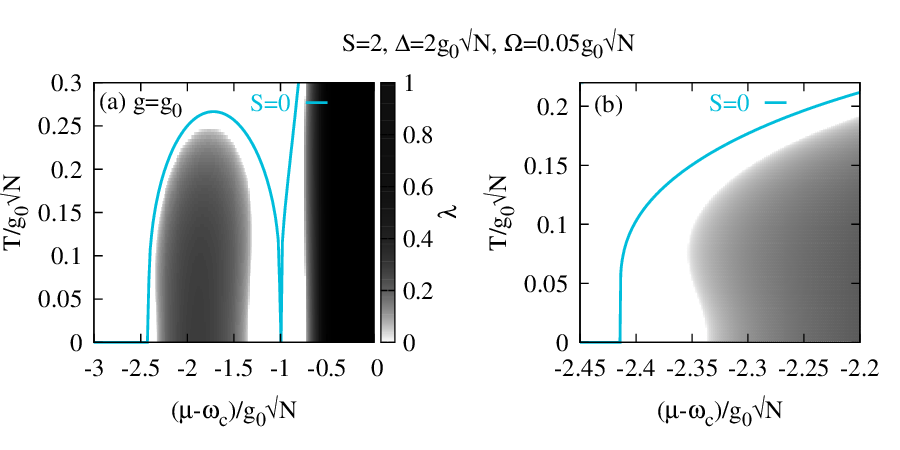}
\caption{(Colour online). (a) Grayscale map of order parameter vs
  temperature and $\mu$, both measured in units of $g_0\sqrt{N}$.
  Solid line shows the phase boundary with $S=0$ for comparison.  (b)
  \ch{A close-up of the re-entrance, shown for clarity.} Parameters:
  $\Delta = \omega_c - \epsilon = 2 g_0\sqrt{N}$ , $S=2$, $\Omega=0.05
  g_0\sqrt{N}$.}
\label{fig:main_phase_diag}
\end{figure}

\ch{Figure~\ref{fig:main_phase_diag} shows a phase diagram: Critical
  temperature as a function of chemical potential}.  \ch{The
  gross structure of the phase diagram seen in
  Fig.~\ref{fig:main_phase_diag}(a) (grayscale) is similar to 
  that seen for $S=0$ (solid blue
  line)~\cite{Eastham2001,Schmidt2013}, having two separate condensed
  regions. Coupling to vibrational modes changes some features. Firstly, the condensed region shrinks. This is because the
  \textit{effective} coupling strength to light is suppressed by the
  dressing of vibrations. Secondly, re-entrant behaviour as a function
  of temperature is introduced, Fig.1(b),
\textit{i.e.}\ on decreasing temperature near the edge of the lobe,
there is a sequence of transitions from normal to condensed and back
to normal. The re-entrance can be explained by the effect of
vibrational sidebands.  Generally, condensation occurs when the
  chemical potential reaches a polariton mode, leading to a
  macroscopic occupation of that mode.}  \ch{If there is a sideband
  below the bare polariton, then condensation will occur at a smaller
  chemical potential.  Such sidebands are associated with transitions
  from a vibrationally excited electronic ground state to an
  electronic excited state with fewer vibrational excitations, as such
  they only occur when vibrational modes are thermally occupied,
  i.e. for, $k_{B} T > \Omega$.  Since the characteristic temperature
  required for the condensed phase (except at $\mu \to \omega_{c}$) is
  $k_{B}T \sim g \sqrt{N}$, re-entrance is only visible if $\Omega \ll
  g\sqrt{N}$.}


\section{Photoluminescence spectrum}
Further understanding of the origin of the re-entrance can be found by
looking at the nature of the mode which condenses.  \ch{For a second
  order phase transition, the mechanism is as described earlier: When
  the chemical potential reaches a bosonic mode, which cannot be saturated, it
  will become macroscopically occupied.}  However, this presents an
apparent paradox for a system with vibrational sidebands.  \ch{The
luminescence spectrum formally has an infinite number of sidebands
both above the bare polariton mode, associated with creating
additional vibrational excitations upon a transition and, at
$T>0$, below the bare polariton, associated with destroying
existing thermally populated vibrational excitations. Thus at any
  non-zero temperature, there are an infinite number of modes below
  the bare polariton, which one would expect to become macroscopically
  occupied, and so condensation would appear to be possible at arbitrarily negative chemical potentials. This is not observed, as shown in
  Fig.~\ref{fig:main_phase_diag}(a).}

\begin{figure}
  \centering
  \includegraphics[width=3.4in]{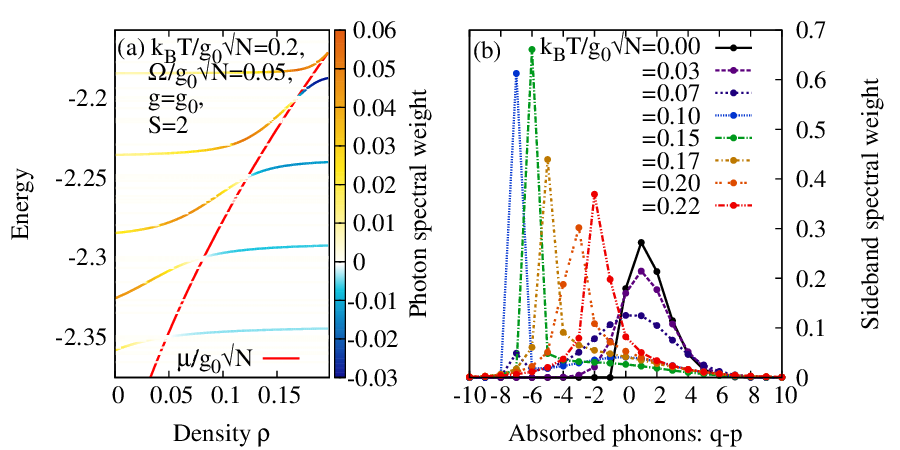}
  \caption{ (Colour online). (a) Spectral weight of normal modes vs
    density of excitations $\rho$ (number of excitons per molecule).
    Lines are coloured according to their spectral weight (white
    corresponding to zero weight).  Solid (red) line shows chemical
    potential.  (b) Spectral composition of the mode which acquires a
    macroscopic occupation for various temperatures.  Each line shows
    the discrete probability distribution for the number of phonons
    absorbed in the associated mode.  Parameters: $S=2, k_B T=0.2
    g_0\sqrt{N}, \Delta = 2 g_0 \sqrt{N}, \Omega = 0.05g_0\sqrt{N}$.}
\label{fig:spectra}
\end{figure}
\ch{The resolution of this apparent paradox lies in the changing nature of the normal modes as one varies the chemical potential.  The normal modes arises from the hybridisation of the photon
with the various vibrational sidebands of the molecular transition.
The photon, being bosonic in nature, has an unbounded occupation. The molecular excitation, on the other hand, is a hard--core boson with  occupation zero or one, and so cannot be macroscopically occupied. In general, polariton modes are superpositions of the bosonic cavity field and the hard--core bosonic molecular excitations. As such, because the nature of the mixture varies, there are points where the mode can be purely excitonic, i.e\ has a vanishing photon component.
Condensation can only occur if the chemical potential hits a mode which has some photon component (measured by the spectral weight).  The crossing of a purely excitonic point would lead to the inversion of two-level systems, but no condensation.} 

\ch{Figure~\ref{fig:spectra}(a) shows the photon spectral weight of the normal modes along with the chemical potential. The spectral weight of the sidebands is increasingly small as one goes to transitions involving
larger differences of numbers of vibrational excitations, but as long
as the spectral weight is not exactly zero, these modes appear
susceptible to become macroscopically occupied.
 There are many places where the chemical potential appears to cross the vibrational sidebands, but crucially, all these crossings
are avoided because the photon spectral weight of the sideband
vanishes at these specific points, \textit{i.e.}\ the mode becomes purely excitonic. After such an avoided crossing, the spectral weight of the mode becomes negative. The Bose occupation function, $n_{B}(\omega)=\left[e^{\beta \left(\omega -\mu \right)} -1 \right]^{-1} $, is negative for energies below the chemical potential. When a negative spectral weight is combined with a negative occupation, they give a meaningful positive luminescence spectrum~\cite{agd} $P(\omega) = -
n_B(\omega) \Im\left[\pi \mathcal{D}(i\omega_n =
  \omega-\mu+i0)\right]$}.
\ch{Technical details of the calculation of the Green's function, $\mathcal{D}$, are given in~\cite{supdat}.}

\ch{Figure~\ref{fig:spectra}(b) shows the composition of the mode which
acquires a macroscopic occupation, exactly at the critical point, for
various temperatures (see~\cite{supdat} for how this is
calculated).} At low temperature the number of vibrational quanta can
only increase; there are no vibrational excitations in the ground
state.   At high enough temperature, $k_{B}T >
\Omega$, the vibrational mode of the electronic ground state is
thermally populated. As a consequence, the re-entrant behaviour as a
function of $T$ appears in Fig.~\ref{fig:main_phase_diag}(a). This
corresponds to the emission of vibrational quanta in the electronic
transition at the critical point, \textit{i.e.} spectral weight occurs
at negative $q-p$, as clearly illustrated in
Fig.~\ref{fig:spectra}(b).


\section{First order transition in large $S$ limit \& variational
  polaron transform}

\begin{figure}
\centering
\includegraphics[width=3.4in]{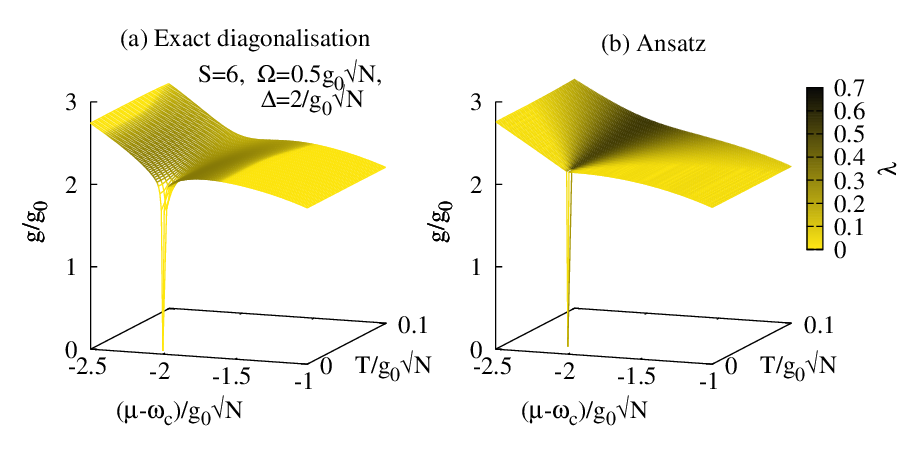}
\caption{(Colour online). Critical $g/g_0$ vs chemical potential and
  temperature for $S=6$ .  \ch{Calculated by (a) exact diagonalisation of Eq.~(\ref{eqn:exactDiag}),
  (b) vibrational MF ansatz.}  The colour scale shows
  $\lambda=\langle \psi \rangle/\sqrt{N}$ at the phase boundary: Light
  (yellow) colours imply $2$nd order transition whereas dark (black)
  strongly $1$st order. Around resonance the transition is always
  (weakly) $1$st order.  Parameters: $\Omega=0.5g_0\sqrt{N}, \Delta
  =2g_0\sqrt{N}$.}
\label{fig:S6_phase}
\end{figure}

We now turn to consider the behaviour when $S \gg 1$. \ch{Such values, which correspond to ultrastrong coupling to vibrational modes, are
interesting because, as seen in Fig.~\ref{fig:S6_phase} for $S=6$,
they can lead to  first order transitions into the superradiant
state.}  At small $T$ the strength of the first order jump is largest
at points near to, but not exactly at, $\mu=\epsilon$.  For yet larger
$S$ (not shown) a first order jump can also occur within the condensed
phase. If viewed as a dynamical system, the first order jump in
  the field could be potentially regarded as an optical switch. \ch{The
  size of jump of $\lambda$ can be controlled by varying the ratio
  $\Omega/g\sqrt{N}$.}  First order transitions have also recently
been noted within other variants of the Tavis--Cummings
model~\cite{Baksic2013}.

The existence of a first order phase transition can be understood by
considering the extent of polaron formation at large $S$,
\textit{i.e.}\ the extent of entanglement of the vibrational and
electronic states of the molecule.  \ch{Since we always consider MF theory
for the photon mode, there is never entanglement between the photon
mode and other modes.  The entanglement we consider corresponds to
conditional displacement of the vibrational mode, dependent on the
state of the two-level system.  Such entanglement reduces the overlap
between the electronic ground and excited states, and so reduces the
coupling to light. Such entanglement also lowers the vibrational
energy.  This is favoured in the normal state.}  In the condensed
state, it is instead favourable to increase the optical polarisation
by suppressing the entanglement, and having similar vibrational
configurations for both electronic states. As a result, at a given
value of $g$ it is possible to sustain a self-consistent solution both
for the normal state and the condensed state.  A first order
transition arises due to switching between these states.


\ch{To further elucidate this first order transition, we introduce a
variational ansatz which captures the behaviour at large $S$.  This
ansatz can be framed as an additional MF approximation for the
vibrational modes.  However, to allow for the entanglement between
vibrational modes and the electronic states discussed above, we
first make a variational polaron
transform~\cite{Silbey1984a,McCutcheon2011c} 
$\hat{H}\to \hat{H}^\prime=e^{\hat{K}}\hat{H}e^{-\hat{K}}$ where
$\hat{K}=\frac{\eta}{2}\sqrt{S}\sigma^{z}(\hat{a}^{\dagger}-\hat{a})$.  This
transform conditionally displaces the vibrational mode dependent on
the state of the two-level system. This  will
therefore transform a product state to an entangled state.  By using such a
displacement followed by a MF theory, \textit{i.e.}\ a product state,
we obtain a state which is effectively entangled in the original
basis, with the entanglement depending on $\eta$.  Thus, this
transform followed by the MF approximation $\hat a \to
\alpha$, corresponds to a variational approach with $\eta$, $\alpha$,
and $\lambda$ as variational parameters.  This ansatz
is valid if the vibrational states are approximately coherent states,
which requires $S \gg 1$.}  The free energy of this vibrational MF
theory can be written as:
\begin{equation}
  \frac{F}{N}=\tilde{\omega}_{c}\lambda ^{2} + \Omega \alpha ^{2} -\frac{\Omega S}{4}\eta \left(2-\eta\right)-\frac{1}{\beta}\ln \left(2\cosh(\beta \zeta)\right)
\end{equation}
where $\zeta =\sqrt{\delta ^{2} +(\tilde{g}\lambda )^{2}}$ is written
in terms of the effective molecular transition frequency $\delta$ and
vibrationally dressed optical coupling $\tilde{g}$, given by:
\begin{equation}
  \delta = \frac{\tilde{\epsilon}+2 \Omega \sqrt{S}(1-\eta)\alpha}{2},
  \qquad
  \tilde{g}=g\sqrt{N} e^{-S\eta ^{2}/2}.
\end{equation}

Minimising this free energy with respect to the variational parameters
$\lambda$, $\alpha$ and $\eta$ we obtain the gap equation
$\tilde{\omega}_{c} \lambda = \tilde{g}^{2} \lambda {\tanh(\beta
  \zeta)}/{2 \zeta}$.
Defining $\kappa = {\tilde{g}^{2}\lambda ^{2}
  \tanh(\beta \zeta)}/[\zeta ^{2}-(\delta \tanh(\beta \zeta))^{2}]$ we
may write the equations for $\alpha, \eta$ as:
\begin{equation}
\alpha =\frac{\delta \sqrt{S} \kappa \tanh(\beta \zeta)}{%
  2 \zeta \kappa +\Omega},
\qquad
\eta = \frac{\Omega}{2 \zeta \kappa +\Omega}.
\end{equation}
The equation for $\eta$, describing the extent of polaron formation,
is instructive.  At small $\lambda$ or large $\Omega$, $\eta \to 1$,
and one has fully developed polarons (fully entangled vibrational and
electronic states).  If, on the other hand, $\lambda$ is large, \textit{i.e.}\
the drive is strong, then $\eta \to 0$ and the polaron formation is
suppressed~\cite{Silbey1984a,McCutcheon2011c}.  In addition to this
behaviour, typical of a variational polaron transform, there is an
extra level of self-consistency here: The photon field $\lambda$
depends on the polarisation of the molecules, and hence the effective
coupling strength $\tilde{g}$.  When the bare coupling $g$
is small, the photon field is small, and polarons are well developed,
further suppressing the effective coupling $\tilde{g}$. At larger $g$,
polaron formation is suppressed, producing a stronger effective
coupling $\tilde{g}$.  At zero temperature, this leads to a jump
within the condensed phase, between a weakly and strongly polarised
phase.  Within the variational approach, such a jump occurs near
$\epsilon=\mu$ if $S > 27/8$.  \ch{At non-zero temperature the same effect
leads to a first order normal to condensed phase transition.}

\ch{Figure~\ref{fig:S6_phase} shows the close match between the
variational polaron transform and the exact diagonalisation of Eq.~(\ref{eqn:exactDiag}) in the
large $S$ limit.}  In this limit, one expects the vibrational state can
be approximately described as a coherent state.  For smaller $S$, the
vibrational MF theory predicts more strongly first order
transitions than the exact diagonalisation. Exact diagonalisation
shows that the first order jump reduces and the transition becomes
second order as $S\to 0$ for all values of $\mu$.  Note that in the
limit $S\to0$, the vibrational modes are irrelevant, and both the
vibrational MF theory and the numerical diagonalisation
reduce to the Tavis--Cummings model.

\section{Conclusion}
In conclusion, we have studied the Tavis--Cummings model dressed by
local vibrational modes.  Coupling to vibrational modes suppresses the
critical temperature for the superradiant phase and can induce
re-entrance, \textit{i.e.} a sequence of normal-condensed-normal transitions as
a temperature decreases.  For sufficiently strong coupling to
vibrational modes, the phase transition can become first order, which
can be understood within a variational polaron ansatz.  \ch{We also
resolved an apparent paradox raised by the existence of vibrational
sidebands in the luminescence spectrum: Although vibrational sidebands
exist below the bare polariton frequency, 
condensation at most of these sidebands is avoided as the photon
spectral weight vanishes when the chemical potential crosses such
modes.}  Nonetheless, at large coupling to vibrational modes, the
polariton mode which does condense contains a strong admixture of
different vibrational sidebands.  These results illustrate the rich
possibilities arising from collective effects when coupling electronic
systems to both radiation and vibrational modes.


\acknowledgments{\ch{ We
  acknowledge discussions with S. Yarlagadda, B. Lovett and R. Gomez-Bombarelli. JAC acknowledges support from EPSRC, SR from the Cambridge Commonwealth trust, JK from EPSRC program ``TOPNES'' (EP/I031014/1) and EPSRC
  (EP/G004714/2).} Argonne National Laboratory's work supported by the
  U.S. Department of Energy, Office of Basic Energy Sciences under
  contract no. DE-AC02-06CH11357. 
  JAC performed the numerical calculations, and SR performed the variational calculations. All authors contributed to the conception of the project, and the preparation of the manuscript.
  }

\bibliographystyle{eplbib2}

\end{document}